\documentclass{elsart}
\usepackage{graphicx,harvard,amssymb}

\def\url#1{\texttt{#1}}
\def\bibcode#1{(\texttt{#1})}

\begin{document}
\begin{frontmatter}

\title{Tidally-Driven Transport in Accretion Disks in Close Binary Systems}

\author{John M. Blondin\thanksref{john}}
\address{Department of Physics, North Carolina State University, Raleigh NC}
\thanks[john]{Email: john\_blondin@ncsu.edu}

\begin{abstract}
{The effects of binary tidal forces on transport within an accretion disk
are studied with a time-dependent hydrodynamical model of a
two-dimensional isothermal accretion disk.  Tidal forces quickly
truncate the accretion disk to radii of order half the average radius
of the Roche lobe, and excite a two-armed spiral wave that remains stationary
in the rotating reference frame of the binary system.
We measure an effective $\alpha$ of order 0.1 near the outer edge of
the disk in all of our models, independent of the mass ratio, Mach number,
and radial density profile.  However, in cold disks with high Mach number,
the effective $\alpha$ drops rapidly with decreasing radius such that 
it falls below our threshold of measurement ($\sim 10^{-3}$) at a radius
of only one third the tidal radius.  In warmer disks where the Mach numbers
remain below 20, we can measure an effective $\alpha$ down to radii
10 times smaller than the maximum size of the disk. }
\end{abstract}


\end{frontmatter}

\section{Introduction}

Accretion disks play a prominent role in modern astrophysics, yet we
remain largely ignorant of their detailed structure and local
dynamics.  The importance of angular momentum transport has been
understood for a long time \cite{sands,lbell}, but
for many years our understanding of disks has been built on a
parameterization of some unknown transport mechanism.  In recent years
much progress has been made in elucidating the role of
magnetohydrodynamical turbulence in mediating angular momentum
transport in accretion disks.  A recent review of our current
understanding of transport is given by \citeasnoun{bandh}.

In addition to local sources of transport such as MHD turbulence, global
processes may also contribute to the transport of angular momentum in accretion
disks.  In particular, spiral shock waves can act as a local sink for
angular momentum \cite{spru87,larson}.  
Because these waves
are traveling slower than the local Keplerian velocity, they contain
negative angular momentum, and any dissipation at the shock will
decrease the angular momentum of the orbiting gas.  While spiral waves
can be produced by a variety of sources, they arise naturally in
binary star systems such as X-ray
binaries, cataclysmic variables, and binary proto-stars,
where the gravitational pull of a binary companion creates a two-armed spiral
shock wave that co-rotates with the binary system.  

Such spiral waves have been observed in several numerical studies,
beginning with the early work of \citeasnoun{sawada} and continuing with
recent attempts to model the observed spiral pattern in IP Peg \cite{gll98,phil}.
Despite this large body of numerical work, there has been relatively
little progress in the quantitative measurement of the effective transport
of spiral shocks in hydrodynamic disks.
This is due in large part to the fact that it is easy
to set up a hydrodynamic simulation of a Keplerian accretion disk, but it is
very hard to do it accurately enough to measure the relatively small
transport expected due to spiral waves.  One needs high numerical accuracy, a
large range in radii, and long simulations covering several orbital periods.
\citeasnoun{rs93} ran a simulation without a tidal stream for tens of orbits.
While their models did exhibit a stationary two-armed spiral pattern, 
the shock dissipation was not enough to keep up with the cooling in their
model.  As the gas in the disk cooled and the Mach number increased,
the spiral shocks became weaker and eventually faded away.  
\citeasnoun{savon} performed careful numerical simulations to study
tidal effects, making a quantitative comparison with analytic calculations
of the linear response of the disk to tidal forces.  They found that
spiral shocks were much diminished in cold disks with Mach numbers
of order 25 or higher, a feature they attributed to a mismatch between
the wavelength of the tidal response of the disk and the length-scale of 
the driving tidal force.
\citeasnoun{godon} ran relatively high resolution simulations using a 
markedly different numerical technique (a hybrid spectral method) and found 
results similar to previous authors, namely a strong two-armed spiral wave.
Note that both \citeasnoun{savon} and \citeasnoun{godon} had disks 
that covered only a limited range in radii (4 and 5 respectively), and
so could not follow the propagation of spiral waves deep into the disk.

The results of these simulations suggest that spiral waves are
a robust feature of accretion disks in binary systems, and that these
spiral shocks can indeed transport mass and angular momentum through the
disk.  They do not, however, provide a quantitative measure of this
transport.  The goal of this paper is to measure an effective $\alpha$
due to spiral shocks excited in accretion disks by the tidal forces 
of a binary companion.

\section{An Idealized Accretion Disk}\label{sec:model}

We begin by defining an idealized theoretical model of an accretion
disk that we can simulate with a standard time-dependent hydrodynamics
code.  Of particular concern is the treatment of the thermal structure
of a disk, which is determined by a balance between local energy
deposition and vertical energy transport via convection and radiative
diffusion.  If an adiabatic equation of state is used and no outlet is
provided to remove the thermal energy, the disk will gradually heat
up.  This approach has been used by several authors \cite{sawada}, but
the increasing sound speed changes the model as the simulation evolves,
and ultimately leads to hot disks with a correspondingly large scale
height and small Mach number.

In order to include the thermal structure of 
the disk, a model must include energy transport in the vertical direction.  
One way to do this without including radiative diffusion and three 
dimensions (a difficult prospect even for the fastest computers)
is to model energy loss by including an ad hoc
cooling function designed to remove energy at the same rate one would
expect due to thermal radiation from the disk surface.
\citeasnoun{rs93} used this approach, specifying a local cooling rate
proportional to $T^4$.  While this approach will not necessarily produce
a realistic temperature structure in the presence of shocks, it does
allow the use of a realistic value of $\gamma$ while keeping the
temperature in the disk well below the local virial temperature. 
\citeasnoun{godon} avoided the issue of shock heating by assuming
a polytropic equation of state,  $P = K\Sigma^\gamma$, 
where $K$ is a predefined constant.
\citeasnoun{savon} took an intermediate approach and included 
an energy equation but added local 
cooling to maintain a polytropic equation of state.

We take a slightly different approach and assume an isothermal equation
of state.  This is equivalent to the polytropic equation of state in
the limit that $\gamma=1$.  The assumption of an isothermal gas
provides for a simpler problem, fewer parameterizations, and allows for
faster computation.  However, the analysis of \citeasnoun{spru87}
suggests that spiral waves will not propagate for a strictly isothermal
equation of state, so we may be underestimating the role of spiral
shocks in this work.

An ideal, isothermal gas evolves according to the Euler equations, 
which can be written in conservative form as:
\begin{equation}
{\partial \rho\over\partial t} + \nabla\cdot (\rho \vec u) = 0
\end{equation}
\begin{equation}
{\partial \rho\vec u\over \partial t} + \nabla\cdot (\rho\vec u\vec u)
+ \nabla P = \rho\vec a,
\end{equation}
where $\rho$ is the mass density, $\vec u$ is the velocity, and
the pressure is given by $P=c_s^2\rho$ with $c_s$ being the 
isothermal speed of sound.  In a coordinate frame
Co-rotating with the binary system, the acceleration of the gas can be
written as
\begin{displaymath}
\vec a = -\nabla\Phi - 2\vec\Omega\times\vec u
\end{displaymath}
where the last term is the Coriolis force due to the Co-rotating 
coordinate system, $\Omega$ is the orbital frequency of the binary system, and 
the effective binary potential is given by
\begin{displaymath}
\Phi = -{GM_1\over r_1} -{GM_2\over r_2} - {1\over 2}(\vec\Omega\times
\vec r_{cm})^2.
\end{displaymath}
The last term is the centrifugal force due to the Co-rotating system,
and $r_1$ is the distance to the accreting star, $r_2$ is the distance
to the donor star, and $r_{cm}$ is the distance to the center of mass of
the binary system.
We have normalized our model such that the unit of distance is 
given by the binary separation and
$GM_1 = 1$.  In these 
normalized units the orbital period is $P = 2\pi/\sqrt{1+q}$, where
$q=M_2/M_1$ is the mass ratio.

We further restrict our idealized model to two dimensions by assuming
no vertical motions nor variations with height in the disk.  When using 
a cylindrical coordinate system, this is equivalent to replacing the 
density in the disk with a surface density, $\Sigma = 2\rho H$, and
assuming a constant scale height, $H$\cite{godon}.

Because our objective is to measure the tidal effects on transport in an
accretion disk in a binary system, we need some quantitative measure of
transport through the disk.  The most direct physical quantity we could
measure is the mass accretion rate, given by $\dot M = 2\pi r\Sigma v_r$
in two dimensions.  The surface density 
has no particular scale (we chose $\Sigma = 1$ in our flat disk models);
the important parameter is really the radial velocity, $v_r$. However,
rather than reporting the average radial velocity (as compared to, say, 
the sound speed), we have followed the historical convention of this 
subject and reported the effective value of $\alpha$. 
\citeasnoun{sands} introduced $\alpha$ as a means of parameterizing some
unknown form of angular momentum transport, presumably related to
local turbulence in some fashion.  While the transport being measured in
our models is a global phenomenon rather than a local process, we can
still refer to an {\it effective} $\alpha$ that would give the same mass
accretion rate in an $\alpha$-disk model that we find in our
simulations.

The standard
$\alpha$-disk theory gives the mass accretion rate in terms of $\alpha$
as $\dot M = 3\pi\alpha c_s H\Sigma$, where $H$ is the local scale height in
the disk.  Using the fact that $H=Rc_s/v_\phi$ for a cold, thin disk, we
find an equation for $\alpha$:
\begin{equation}
\alpha = {2\over 3}{v_\phi \langle v_r\rangle\over c_s^2},
\end{equation}
where $\langle v_r \rangle$ is a density-weighted average of the radial velocity:
\begin{displaymath}
\langle v_r\rangle = {\int_0^{2\pi}\Sigma v_r d\phi \over
         \int_0^{2\pi}\Sigma  d\phi }.
\end{displaymath}
Thus, to measure $\alpha$ we need to keep track of the local radial
mass flux ($ v_r\Sigma$) and the local surface density.

It is important to point out that our ideal accretion disk is not
expected to be in strict steady state.  We have no a priori reason to
expect the mass accretion rate to be independent of radius.  Mass
accretion in this model is driven solely by the action of spiral shock
waves, whose strength will certainly depend on the radius within the
disk.  The propagation of these shocks will also depend on the density
profile in the disk, but there is no guarantee that a combination of
density profile and shock propagation exists that leads to a constant
mass accretion rate.  Furthermore, even if there did exist a
steady-state solution, we may not, in practice, be able to evolve a
simulation long enough to reach a true steady state.

\section{Computational Method}\label{sec:method}

Our idealized accretion disk model is evolved using a modified,
isothermal version of the hydrodynamics code 
VH-1\url{http://wonka.physics.ncsu.edu/pub/VH-1}, run in
two-dimensional cylindrical coordinates ($r$, $\phi$) centered on 
the accreting star.  This code is
based on the Lagrange-remap version of the piece-wise parabolic
method \cite{candw}.  The isothermal version is significantly faster
than the adiabatic code because there is no need to solve an energy
equation and there exists an analytic solution to the Riemann problem,
avoiding a costly iterative solution as required in an adiabatic code.

There are essentially two modifications to the standard distribution of
the isothermal version of VH-1: Forces to account for the rotating binary
potential and changes in the remap step to allow conservation of angular
momentum.  This second step is vital to any quantitative study of 
accretion disks, for it is the transport of angular momentum that allows
accretion to take place.  If a numerical method does not conserve angular
momentum, accretion may result solely from, or be completely wiped
out by, numerical error.

Conserving angular momentum in VH-1 is a fairly simple task.  When
computing the hydrodynamic evolution in the $\phi$ direction the radius
is held constant so conserving angular momentum is equivalent to
conserving $\phi$ momentum, which VH-1 already does.  When computing
the evolution in the radial direction the angular momentum in a zone is
left unchanged during the Lagrangian hydrodynamic evolution.  This
leaves the remap step in the radial direction as the only place in the
standard code where angular momentum is not conserved.  Fortunately,
conserving angular momentum in this step is trivial: instead of
interpolating on $v_\phi$ and remapping $\phi$ momentum, one need only
interpolate on $rv_\phi$ and remap angular momentum.  Doing so will
conserve the total angular momentum on the grid to within machine
roundoff error (if boundaries are reflecting and forces are only
radial).

An additional concern is the local diffusion of angular momentum.  Even
if angular momentum is conserved globally, it may diffuse across the
numerical grid, producing anomalous transport.  \citeasnoun{savon}
modified their code to interpolate on the quantity $\Sigma
r^{1/2}v_\phi$ (which is flat in a Keplerian disk with constant
$\Sigma$) in order to minimize the diffusion of angular momentum.  This
prevented unwanted instabilities from developing in their disk,
although it did not necessarily guarantee conservation of angular
momentum.   Fortunately, VH-1 exhibits very little diffusion of angular
momentum in this problem, a feature we attribute to the higher-order
interpolations used in the piece-wise parabolic method.

The Lagrange-remap method used in VH-1 allows a simple, accurate
measure of the radial mass flux need to compute an effective $\alpha$.
Advection in VH-1 is treated in the remap step, where gas that has been
evolved on a Lagrangian grid is remapped back onto a fixed Eulerian
grid.  For example, if the Lagrangian interface between zones 1 and 2
moves to the right during a timestep, the remap step takes the mass
(and momentum) within a sub-shell defined by the original zone
interface and the final Lagrangian interface and instantaneously moves
it from zone 1 (where this mass was before the timestep) to zone 2
(where it has advected during the timestep).  The mass flux through the
Eulerian interface during the timestep is just the mass in the
sub-shell that was remapped from zone 1 to zone 2.  So, to compute an
average mass flux, $v_r\Sigma$, we summed up all the mass being
remapped across a given radial boundary (i.e., for all values of
$\phi$) during a prescribed interval of time (typically 50 time
steps).  Dividing by the average (over time and $\phi$) density in
these sub-shells, we can calculate an effective value of $\alpha$.

The initial conditions for the disk simulations include a low-density hole
in the first 5 radial zones ($10^{-3}$ of the disk density) to act as a 
buffer between the inner edge of the disk and the inner boundary, zero radial
velocity, and $v_\phi = r^{-1/2} - \Omega r$.  The last term in $v_\phi$ 
accounts for the centrifugal force in the rotating reference frame, and 
provides for an initial disk much closer to equilibrium than pure
Keplerian rotation.

The $\phi$ boundaries are periodic and the radial boundaries are
constructed to allow material to flow off the grid, but not onto the
grid.  At the inner radial boundary the $\phi$ velocity is set to zero
in the ghost zones (extra zones used for interpolation
on the real grid), creating an
imbalance of forces at the inner edge of the hydrodynamic grid
that drives quick accretion off the inner edge.  
Contrary to some previous numerical work
\cite{mats90}, the exact boundary
conditions at the inner edge do not significantly alter the simulations.
Even with a reflecting boundary, material piles up at the inner edge of
the grid without influencing the spiral shocks at larger radii.
The outer boundary also does not exert much influence on the long
term evolution of the disk.  Once tidal forces trim down the outer edge
of the disk, the density in the outer zones is very small so there is 
negligible flux of mass and momentum off the outer edge of the grid.

Most of our simulations were run on grids with a radial span of
a factor of 20, with the radial zone size gradually increasing with
increasing radius to provide consistent resolution over the full range
of radii in the simulation.  The outer radius was chosen to be approximately 
the distance to the inner Lagrange point, allowing plenty of room
between the outer edge of the tidally-truncated disk and the outer 
grid boundary.  One would like to make the inner edge of the numerical
grid as small as possible, but restrictions on the timestep in an
explicit hydrodynamic code make small inner radii very expensive 
to compute.  For stability, the timestep must be less than $r\Delta\phi
/ v_\phi$.  But $v_\phi = r^{-1/2}$ in an accretion disk, so the
limiting timestep decreases as $r^{3/2}$.  Evolving a disk that 
extends down a factor of 20 in radius thus requires a small timestep and
a corresponding large number of cycles; some of our simulations ran
for over $10^6$ cycles.  The choice of 20 for the radial span was based
on the radial decay of $\alpha$ in our standard model, 
as we will see in the following section.  This inner radius also
corresponds roughly to the point of closest approach for a ballistic stream of
gas originating at the $L_1$ point in the binary system.  Since our goal
in future work is to quantify the effects of the tidal stream on
transport within the disk, we have extended our disks in this work down
to this small radius.

\begin{figure} 
\includegraphics[width=350pt]{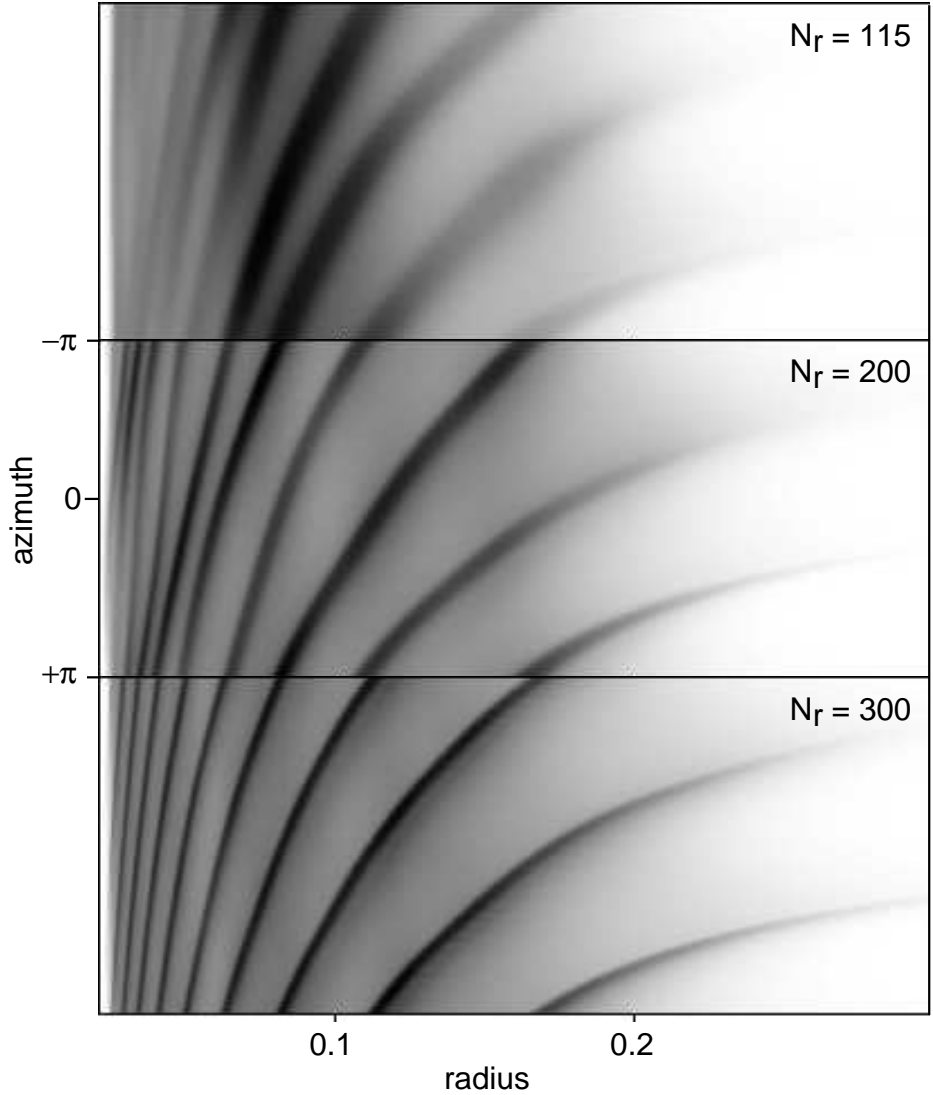}
\caption{Resolution of spiral waves as a function of the number of
radial zones, $n_r$.  The grey scale represents gas density, 
which reaches a maximum of approximately 3 times the initial 
disk density in the spiral waves.} 
\label{fig:resolve} 
\end{figure}

The issue of numerical resolution was approached empirically.  Keeping
in mind that our goal was a quantitative measure of transport effects
caused by tidally-induced spiral waves, we ran several simulations with
different numerical resolutions.  We found relatively little difference
in the density structure of the disk and the computed value of $\alpha$
when we used 115 zones in the $\phi$ direction compared with 200 zones,
so we opted for the former in most simulations.  For the radial
direction we found significant differences between runs with 115 and 200
radial zones, but relatively little difference between runs with 200 and
300 radial zones.  Figure \ref{fig:resolve} shows the loss of definition
in the spiral waves when the radial resolution is too low.  The computed
value of $\alpha$ in the run with 200 zones evolved very similar to that
in the run with 300 zones, but the run with 100 zones looked completely
different (and in fact was negative throughout the evolution).  More
radial zones were required for the colder disk simulations described
in Section \ref{sec:mach}

Finally, a small amount of dissipation was included in the code in the
form of wiggling the grid back and forth a small amount in the radial
direction \cite{candw}.  This was necessary in order to suppress two
different numerical instabilities found in these disk simulations.  The
first creates a zone to zone striping in the radial direction, and the
second leads to epicyclic waves, usually (but not always) appearing at
the inner boundary of the disk.  The first problem is related to the
conservation of angular momentum.  In a one-dimensional (radial) problem
using Lagrangian coordinates, if a given zone has too much angular momentum
it will drift outwards.  However, this motion will drag out neighboring
zones, which will find themselves having too little angular momentum
for their new radius and will drift inwards.  This can lead to a 
striping effect with the radial velocity in each zone changing
sign from zone to zone.
We were not able to fully understand the second problem, but we found
that it was worse (in the absence of grid wiggling) if the ratio of
the sound crossing time across a radial zone to the orbital period at
that radius was larger than some critical value.
For all practical purposes these
instabilities were eliminated by the grid wiggling, which in turn had
negligible effect on the structure of the disk or the computed value of
$\alpha$.

As a test of our computational method we ran a simulation without the
binary potential, i.e., with only the central force of the accreting
star's gravity.  We instituted reflecting boundary conditions at both the
inner and outer radii, so the total mass and angular momentum on the grid
should remain fixed.  To make the test meaningful, we perturbed $v_\phi$ in
a small band of radii in the middle of the grid to create wave
motions.  The mass and angular momentum did remain constant to within
machine round off error.  The local radial velocities remained less than
$\sim 10$\% of the sound speed, and the computed average of $\alpha$
was $\sim 10^{-3}$. This then represents the limit to which
we can measure an effective $\alpha$ in our binary simulations.

\section{Numerical Simulations}\label{sec:results}

This section describes seven hydrodynamic simulations varying the 
sound speed, mass ratio, and density profile in the disk.  The
parameters describing each of these models are listed in Table 
\ref{table:parameters}.

\begin{table}
\caption{Input parameters (sound speed $c_s$, mass ratio $q$,
and surface density index $n$) for the hydrodynamic simulations.
The tidal radius, $R_T$, and the Mach number of the Keplerian
flow at half the tidal radius are given, as well as the ratio 
of the tidal radius to the radius of the Roche lobe
of the accreting star.}
\begin{tabular}{l l r l c c l}
\hline
Model    & $c_s$ & $q$ & $n$ & $R_T$ & $R_T/R_L$ & $M(R_T/2)$\\
\hline
standard & 0.25   & 1.0 & 0 & .214 & 0.56 & 12.2\\
colder   & 0.125  & 1.0 & 0 & .264 & 0.69 & 22.0\\
coldest  & 0.0625 & 1.0 & 0 & .300 & 0.79 & 41.3\\
steeper  & 0.25   & 1.0 & 1 & *    &  *   & 12.2\\
steepest & 0.25   & 1.0 & 3 & *    &  *   & 12.2\\
lowmass  & 0.25   & 0.2 & 0 & .307 & 0.59 & 10.2\\
highmass & 0.25   & 5.0 & 0 & .185 & 0.74 & 13.2\\
\hline
\end{tabular}
\label{table:parameters}
\end{table}

\subsection{Standard Model}

For the purposes of presenting our results we have adopted a standard
model in which $q=1$ and $c_s=0.25$.  We will describe this model in
detail, and then go on to discuss how other models differ from this
standard.  With the adopted sound speed, the Mach number of the disk
material at the inner edge of the disk is $\sim 28$, comparable to some
of the best previous numerical work, but still lower than one might
expect in real disks.  Tidal forces quickly truncate the disk at an
outer radius of $0.21$, which is roughly half of the averaged Roche
lobe radius as given by the analytic expression of \citeasnoun{eggl}.
This disk is significantly smaller than predicted by theoretical
arguments for infinitely cold disks\cite{paczyn}, but one expects that
finite pressure effects will increase the effectiveness of tidal forces
in truncating the disk.  At this tidally-truncated outer edge the disk
material orbits with a Mach number of $\sim 8$.

\begin{figure} 
\includegraphics[width=400pt]{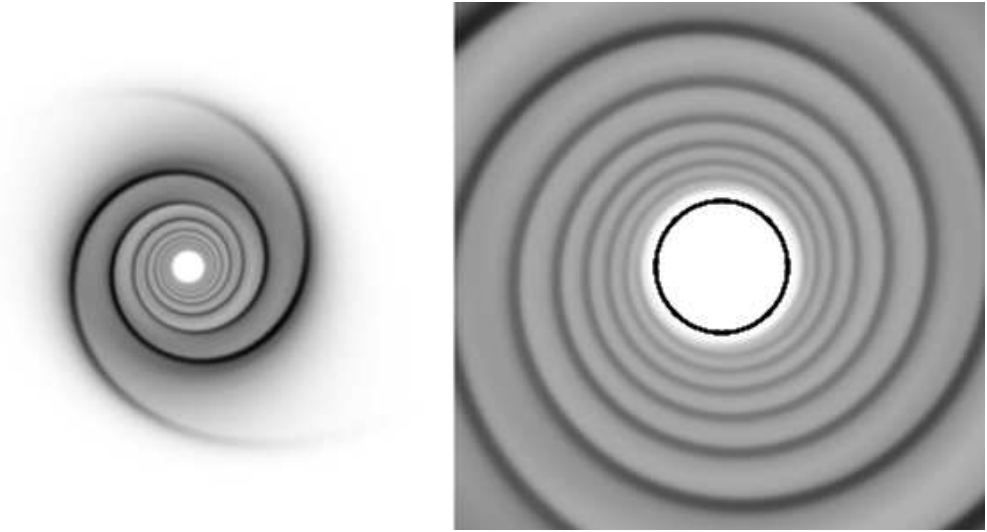} 
\caption{Density in the standard model at time $t=9$.  The left image
shows the entire grid spanning the Roche lobe of the accreting star, 
while the right image shows the central region magnified
by a factor of 5.  The binary companion is to the right of the image,and
the disk orbits in a clockwise sense.
Black corresponds to high density gas, with 
saturation occurring at a density of 3.  The black circle in the right image
marks the inner edge of the numerical grid at $r=0.02$.  The accompanying
animations show the evolution of the disk from the same two perspectives.}
\label{fig:standard} 
\end{figure}

The images and associated animations in Figure \ref{fig:standard} show
the strong spiral structure that develops in our standard model.  The
two-armed spiral wave remains relatively fixed in the co-rotating
reference frame of the hydrodynamic simulation.  The shape of the
spiral is determined by the inward radial propagation of the wave at a
velocity of order the sound speed and the shearing of this wave by the
orbital motion.  The result is a trailing angle (between the spiral
shock and a circle of given radius) given by $\tan\Theta \approx
(c_s/\Omega r)$ \cite{gll98}, or the inverse of the Mach number of the
orbital motion, $M$.  \citeasnoun{larson} suggested a radial wave speed
of $\sim 1.5c_s$, whereas our simulations show a spiral shape
corresponding to a wave speed of $\sim 1.3c_s$.  The density jump at
the spiral shocks varies from a maximum of $\sim 2.1$ near the outer
edge of the disk down to a value of $\sim 1.4$ near the inner edge.
The corresponding shock Mach numbers of order $\sim 1.3$ (given by the
square root of the compression) are in agreement with the low Mach
numbers predicted by \citeasnoun{spru87} and with the predicted shape
of the spiral pattern.  Presumably the strength of the spiral shocks
would continue to decrease if we had extended our simulation down to
even smaller radii, as the spiral shock becomes wound tighter and
tighter with increasing orbital velocity.

\begin{figure} 
\includegraphics{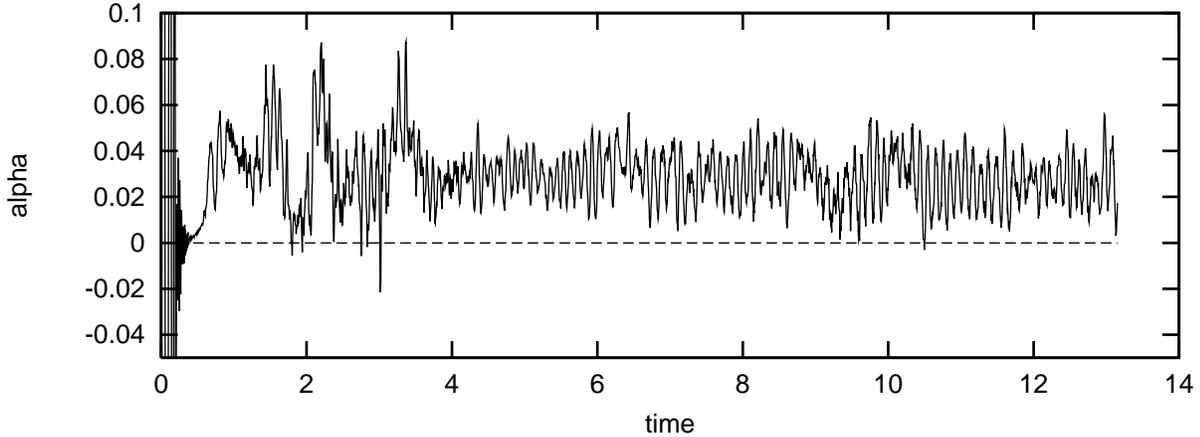} 
\caption{Time evolution of $\alpha$ in the standard model at a 
radius of $r=0.065$. The Keplerian 
period at this radius is $P_{kep}\approx 0.105$.}
\label{fig:alpha_evolve} 
\end{figure}

The initial response of the disk to the binary potential and the
subsequent quasi-steady transport are illustrated in Figure
\ref{fig:alpha_evolve}, which plots the effective $\alpha$ at a radius
of $r\approx 0.065$.  There are very large waves generated by our crude
initial conditions, as shown by the large variations in $\alpha$
shortly after $t=0$, but these quickly dissappear on a timescale of
less than one rotation of the outer disk edge.  The behavior of
$\alpha$ does not appear to settle into a quasi-steady behavior until
after approximately one orbital period (4.44 in this model), but from 1
to 3 orbital periods the behavior appears unchanged.  The high frequency 
oscillations clearly visible in Figure \ref{fig:alpha_evolve} correspond
to the local Keplerian frequency ($P_{kep} = 2\pi r^{3/2} \approx 0.105$).

\begin{figure} 
\includegraphics{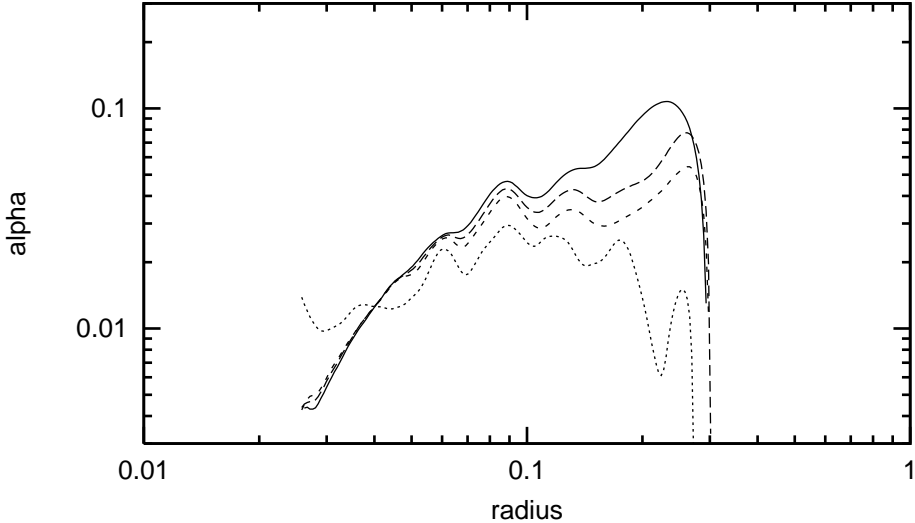}
\caption{Time-averaged radial profile of $\alpha$ 
in the standard model for times  of 7-9 (solid), 9-11 (long dash),
and 11-13 (short dash). The dotted line shows $\alpha$ at late 
times ($t=25-27$) after other wave modes have appeared in the outer
regions of the disk.} 
\label{fig:alpha_prof} 
\end{figure}

The variation of $\alpha$ with radius is shown in Figure \ref{fig:alpha_prof}.
The value of $\alpha$ is as high as 0.1 near the outer edge of the
disk where strong spiral shocks are driven by tidal forces, but appears to
gradually decline with time at large radii.  Inside of $r \approx 0.1$,
$\alpha$ appears relatively constant with time but
drops with decreasing radius.  Although $\alpha (r)$ is not well
fit by a power-law, we can approximate $\alpha$ in the
central region of the disk with $\alpha\approx r^{1.4}$.  At late times
other wave modes appear (see the animation in Fig. \ref{fig:standard}),
slightly reducing $\alpha$ at large radii and enhancing it at small
radii.  The result is a more uniform value of $\alpha$ throughout 
the disk.  However, more analysis of these other wave modes is needed
before one can be confident of their role in angular momentum transport.
For this reason we concentrate our attention on the two-armed spiral waves.

These values of $\alpha$ due to the two-armed spiral shocks
are consistent with the shock strengths quoted
above, but the radial dependence is much steeper than the
predictions of \citeasnoun{larson}, and the absolute values are much
higher than the analytic predictions.  For a spiral shock with Mach
number $M_s$, \citeasnoun{larson} derives an effective $\alpha$ in an
isothermal disk of
\begin{equation}
\alpha \approx 0.07 (M_s^2 - 1)^3\, .
\end{equation}
For an isothermal shock, $M_s^2$ is equal to the compression ratio.
The range in $M_s^2$ from 1.4 to 2.15 gives a range in $\alpha$ of
0.005 to 0.1, consistent with the data plotted in Figure \ref{fig:alpha_prof}.

\citeasnoun{larson} goes on to predict that the strength of tightly
wound spiral shocks in a cold, thin disk should be proportional to
$(\cos\theta )^{3/2}$, where $\theta$ is the trailing angle of the spiral
wave.  Estimating $\cos\theta \approx 1.5c_s/v_o$,
\citeasnoun{larson} finds an effective $\alpha$ of
\begin{equation}
\alpha \approx 0.026 \left( {c_s\over v_o}\right)^{3/2}\, .
\end{equation}
The prediction for an isothermal disk with $c_s=0.25$ is $\alpha
\approx 3.2\times 10^{-3}\, r^{3/4}$, orders of magnitude below that
found in our simulations.

\begin{figure} 
\includegraphics{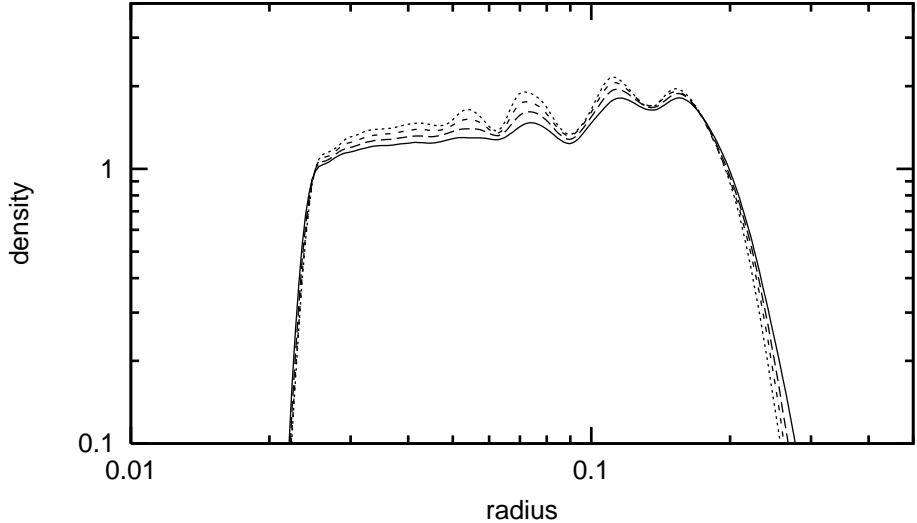}
\caption{Radial profile of $\Sigma$ (averaged over $\phi$)
in the standard model at various times beginning with
$t= 9$ (solid line) and ending with $t=16$ (dotted line). } 
\label{fig:dens_prof} 
\end{figure}

A simple estimate of the timescale for gas to move through the disk 
is $r/v_r$, which for this model ranges from 60 at the outer edge
of the disk to $\sim 300$ near the inner edge.  Having evolved our
simulation for a time of only 13, we do not expect our model to 
have settled into an equilibrium.  We can in fact see small changes 
in the density profile of the disk as plotted in Figure \ref{fig:dens_prof}.
Despite the fact that the density is slowly increasing over most of the
disk, the total mass on the computational grid has decreased by
almost 1\% during the time from $t=9$ to $t=16$.

The two-armed spiral pattern seen in Figure \ref{fig:standard} is not
the only wave present in the disk.  On a timescale of several binary orbits
other wave modes begin to appear.  By viewing the animation accompanying
Figure \ref{fig:standard}, one can identify a purely radial wave ($m=0$)
and a precessing eccentric wave ($m=1$) growing in the vicinity of
the outer disk edge.  While these waves continue to grow to some apparent
saturation level, they do not completely disrupt the two-armed spiral
shock, and $\alpha$ remains at a significant level \ref{fig:alpha_prof}.
Similar wave modes have been seen in the simulations by
\citeasnoun{godon}.  We will discuss them further in the next section,
for they are more apparent in colder disks.

\subsection{Dependence on Mach Number}
\label{sec:mach}

The sound speed in our standard model was chosen to produce 
a significant value of $\alpha$, not because it was astrophysically
interesting.  In most accretion-powered binary systems the Mach 
number of the orbital motion in the disk will be much higher than
we have assumed here.  A colder disk is both more challenging to
simulate and more difficult to measure the expected smaller $\alpha$.
The effect of spiral shocks on transport through an accretion disk
diminishes rapidly with increasing Mach number because the spiral
shocks are wound tighter, making the relative shock Mach number
smaller, leading to less dissipation and effective transport.

In addition to the challenge of measuring small effects, we were
faced with several numerical problems when simulating colder disks.
The first problem is one of resolution.  Because the spiral arms
are wound tighter in colder disks, we found that to accurately
resolve the oblique shocks of the spiral arms, we needed higher
resolution in the radial direction.  
Another difficulty plaguing our cold disk simulations was the increase in
numerical "noise" with higher Mach number.  In one-dimensional
(radial) simulations of Keplerian disks excited with radial waves,
we found a monotonic increase in the average radial velocity with
increasing Mach number in the disk.  While the effective $\alpha$
from this numerical noise remained low ($\alpha = 4\times 10^{-3}$
for $c_s=.0125$), it could affect the measurement of transport 
in very cold disks.  Finally, the numerical instabilities described
in the appendix are more apparent at higher Mach numbers.

\begin{figure} 
\includegraphics[width=400pt]{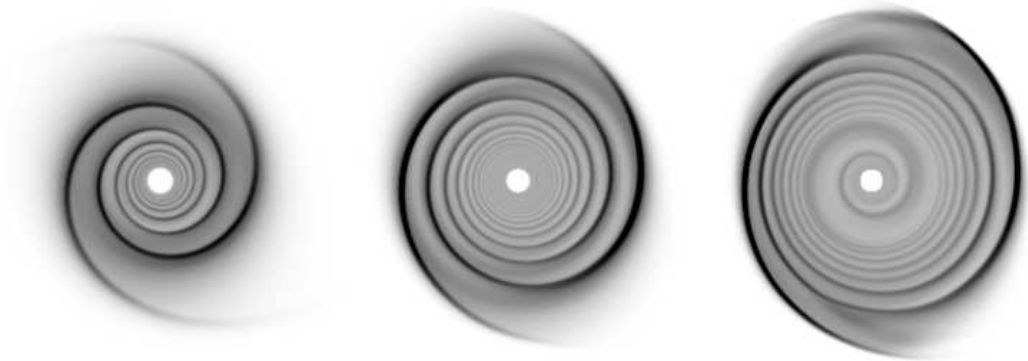}
\caption{Snapshots of the disk surface density in models with a 
sound speed of $c_s = 0.25$ (left), 0.125 (center), 0.0625 (right).
The accompanying animations illustrate the increasingly unsteady
behavior as the sound speed is reduced.} 
\label{fig:mach_den} 
\end{figure}

To investigate the influence of sound speed on our idealized accretion
disks we ran simulations similar to our standard model but with sound
speeds of 0.125 (with 300 radial zones) and 0.0625 (with 400 radial zones).  
As expected, the tidal forces have less of
an effect on the colder disks.  As seen in Figure \ref{fig:mach_den},
the spiral density waves are wound tighter and fade faster with
decreasing radii in the colder, higher Mach number disks.  The two-armed
spiral has almost completely disappeared by the inner edge of the disk
with $c_s=0.125$, and is present only beyond $r\approx 0.08$ in the coldest
disk.

A one-armed spiral is clearly visible in the interior of the coldest
disk.  This wave originates in the outer regions of the disk and
propagates inward, decaying away below a radius of $r\approx .04$.  
There is no trace of this wave at the inner edge of the disk at any
time during the simulation.  A
similar wave can be seen in most of the other simulations, showing up
either as an initial transient, or as a growing mode at late times in
the simulation.  However, this wave is much harder to identify in the
other models, both because the single spiral is weaker,
and also because the double spiral is much stronger in warmer disks and
hence obscures the single spiral.  The one-armed spiral is easily seen
in the animation of the cold disk with $c_s = .125$, but it is a
transient phenomena, presumably excited by the initial conditions.  At
later times there is a faint hint of the one-armed spiral, but it is
much weaker than the double spiral wave.  

The source of this single spiral wave (ignoring for the moment any
transient created by the initial conditions) may be related to the fact that
the double spiral wave at the outer edge of the disk is not steady, and
in fact is much less steady in the coldest disk where the single 
spiral is most prevalent.  In contrast to the 
standard model, there appears to be at least two competing waves excited
at the tidal radius in the coldest disk model, for the two-armed spiral
is never steady.

\citeasnoun{godon} also observed a single spiral in his Model 1, which
has Mach numbers comparable to our coldest disk.  He
attributed this wave to ``viscous oscillations'' excited by the tidal
potential.  In contrast to our simulations however, \citeasnoun{godon}
found that the single spiral wave was trapped between the reflecting
inner boundary and the outer edge of the disk.  As a consequence, this
wave grew unbounded in his simulation and eventually dominated the
entire disk.  In addition, the presence of this wave was closely tied
to his choice of a Keplerian, reflecting inner boundary.  Any deviation
from this assumption and the single spiral did not appear.  In our
simulations, which covered a radial range four times larger, the single
spiral decayed away before reaching the inner edge of the disk, and its
presence was unaffected by the choice of inner boundary conditions.

These simulations also provide evidence that tidal
truncation becomes less effective with increasing Mach number.  The
tidal radius of the coldest disk is $\sim 80\%$ of the averaged Roche
lobe radius, more in line with the analytic predictions of \citeasnoun{paczyn}.
By searching for the largest non-intersecting periodic orbits, 
\citeasnoun{paczyn} found an upper limit (in the absence of pressure
effects) to the size of an accretion disk.  For a mass ratio of unity,
he predicted a maximum disk radius of 0.3, in good agreement with the
tidal radius in our coldest model.  The warmest model had a tidal radius
of 0.214, some 30\% smaller, leaving the disk radius only 60\% of the
Roche lobe radius.

\begin{figure} 
\includegraphics{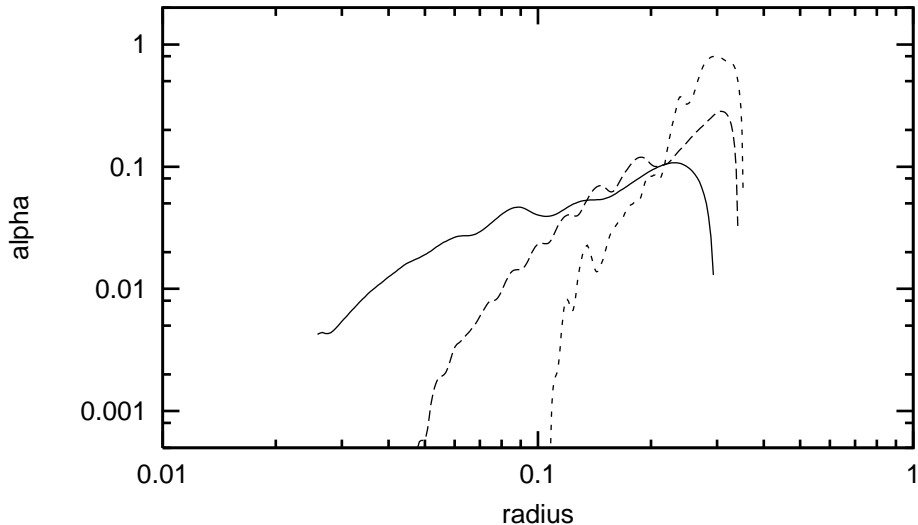}
\caption{Dependence of the effective $\alpha$ on the sound 
speed in the disk. The solid line corresponds to the standard
model with $c_s = 0.25$, the long-dashed line to the model
with $c_s = 0.125$, and the short-dashed line to the model
with $c_s = 0.0625$.} 
\label{fig:mach_alpha} 
\end{figure}

Figure \ref{fig:mach_alpha} shows the precipitous drop in $\alpha$ as a
function of radius for cold disks.  Analytic arguments suggest that
$\alpha$ should drop as $c_s^{3/2}$, but maintain the same $r^{3/4}$
radial dependence in an isothermal disk.  We find that the maximum
value of $\alpha$, found near the outer edge of the disk, remains
roughly independent of sound speed, while the radial dependence of
$\alpha (r)$ steepens with decreasing sound speed.  In fact, because
the radial decay of $\alpha$ is so steep, we found steady mass
accretion only for radii above $r\approx 0.1$ in the coldest disk with
an outer Mach number of $\sim 32$.

\subsection{Dependence on Density Profile}

The results of our models so far are characterized by a relatively
steady mass accretion rate that varies with radius in the disk.  This
implies that the density profile is changing with time as mass piles up
in the disk.  Left to evolve for  a much longer time, we would expect
the density profile to steepen, with higher densities at smaller radii
where the spiral waves are less effective at transporting angular
momentum.  Would this different density profile affect the computed
value of $\alpha$ in the model?  Will the spiral waves weaken as they 
propagate down into regions of higher density?

\begin{figure} 
\includegraphics{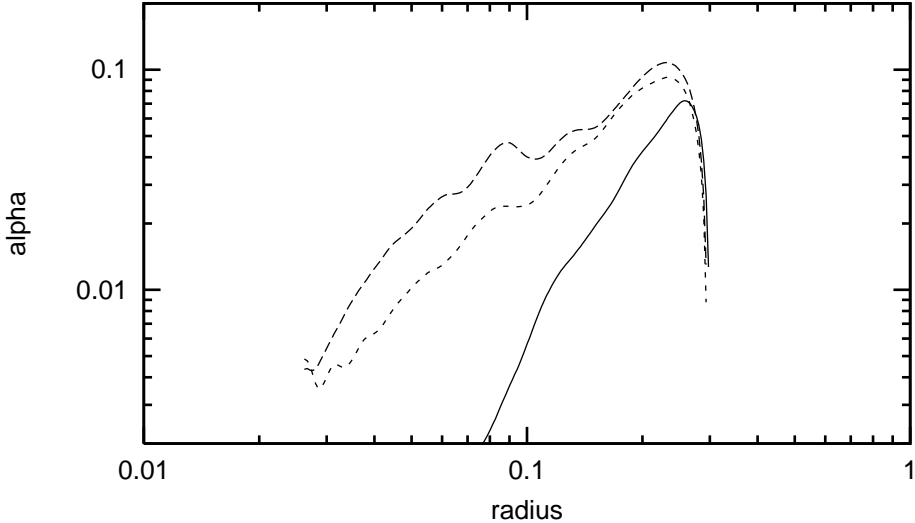}
\caption{Dependence of the effective $\alpha$ on the density
profile in the disk. The solid line corresponds to the 
model with $n = 3$, the short-dashed line to the model
with $n = 1$, and the short-dashed line to the standard model with $n = 0$.} 
\label{fig:steep_alpha} 
\end{figure}

We ran additional simulations for several density profiles $\Sigma\propto
r^{-n}$, and show
results for $n = 0$, 1, and 3 in Figure \ref{fig:steep_alpha}.  
For comparison, an $\alpha$-disk
model for an isothermal equation of state yields $n=1.5$, and for an
adiabatic gas $n=0.75$ \cite{sands}.  Note however that the gas density
scales with a steeper function of radius than the surface density.
In a three-dimensional isothermal model the scale height increases as $H\propto r^{3/2}$,
so the density in the equatorial plane varies as $\rho\propto r^{-3}$.

In all but the most extreme model, the effective transport within the
disk is very similar.  The value of $\alpha$ is down by at most a factor of
$\sim 2$ in the inner disk for the model with $n=1$ as compared to the
standard model, but the radial dependence, maximum values, and tidal
truncation all appear similar in the models with $0 < n < 1.5$.  In
contrast, the spiral waves decay much faster in the steepest disk with
$n=3$.  The spiral waves still extend all the way to the inner edge of
the disk in this model, but the radial velocity perturbations and 
density compressions associated with these waves are lower than found
in the other models with smaller $n$.  As a consequence of this faster
decay, the mass accretion rate is still an increasing function with
radius, despite the fact that the density varies as $\Sigma \propto r^{-3}$.
This result suggests that a true steady state may not exist for a
two-dimensional isothermal accretion disk in which transport is mediated
solely by tidally-induced spiral shocks.

\subsection{Dependence on Mass Ratio}

\begin{figure} 
\includegraphics[width=400pt]{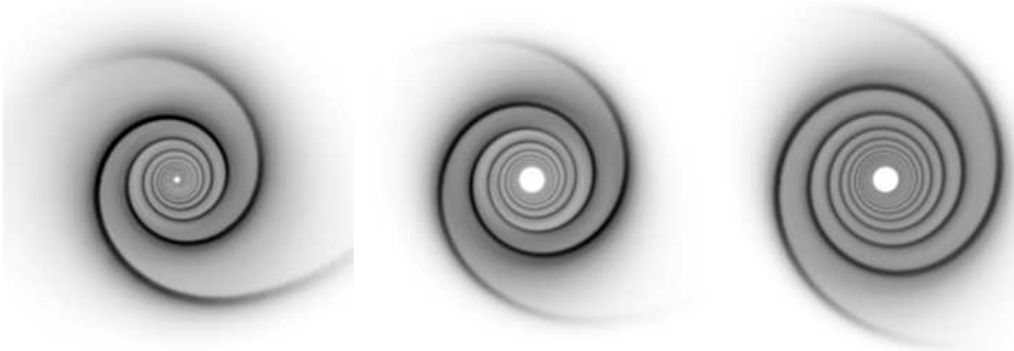}
\caption{Effect of mass ratio, $q$, on the spiral pattern,
with $q=0.2$ on the left, $q=1$ in the center,
and $q = 5$ on the right.  Each image has been scaled to the 
size of the Roche lobe.  The $q=0.2$ model was evolved
on a grid that went down to much smaller radii.  The density scale
in each image has been normalized to the average density in the middle
of the disk.} 
\label{fig:q_den} 
\end{figure}

Our last parameterization is the mass ratio, $q$.  In addition to the
equal mass system studied in the previous subsections, we have run
simulations with a mass ratio smaller than unity (accretion {\it from} a
lower mass companion) and larger than unity (accretion {\it from} a
higher mass companion).  The spiral patterns shown in Figure 
\ref{fig:q_den} are remarkably similar for different values of $q$.
The spiral waves are slightly more open for small $q$.  Since 
the disk extends out to larger radii where the Keplerian velocity
is smaller, the Mach number is lower and the trailing angle of
the spiral is larger.  The strength of spiral shocks, as measured
by the shock compression, is comparable in all three simulations.

\begin{figure} 
\includegraphics{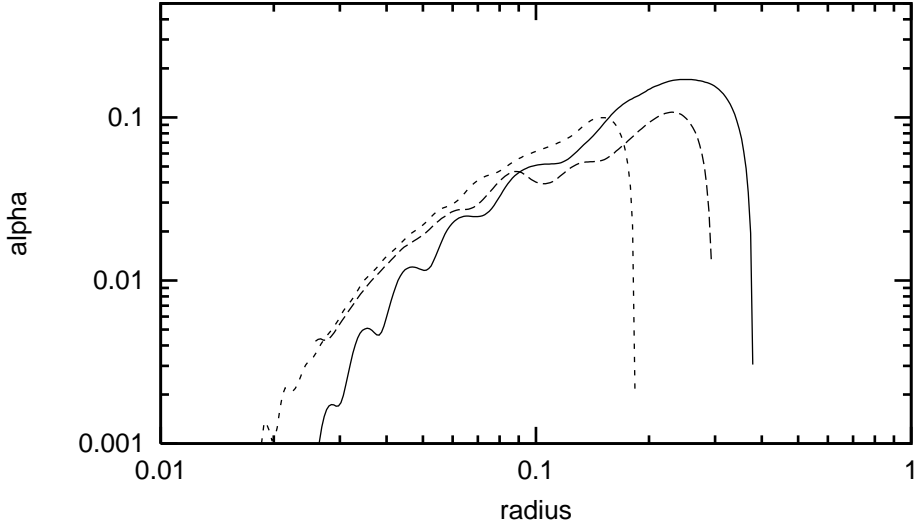}
\caption{Dependence of the effective $\alpha$ on the mass ratio, $q$.
The solid line corresponds to $q=0.2$, 
the long-dashed line to the standard model with $q=1$,
and the short-dashed line to $q = 5$.} 
\label{fig:q_alpha} 
\end{figure}

The relative independence of our model on the mass ratio (at least within
the limited range that we have simulated) is also seen in Figure
\ref{fig:q_alpha}, which shows an effective $\alpha$ that does not vary much
with $q$.  The only significant difference is the location of the 
outer edge of the disk.  But when the tidal truncation radius is compared 
with the radius of the Roche lobe \cite{eggl}, $R_L$, even the size of the
disk does not change much.  Accretion from a low-mass companion produces
a larger disk, but the Roche lobe is larger by almost the same amount,
leading to a ratio of $R_T/R_L$ comparable to a disk in a binary with
equal mass stars.  Accretion from a high-mass companion produces a smaller
disk, but one that occupies a larger fraction of the Roche lobe
(Table \ref{table:parameters}).

\section{Discussion}

The time-dependent hydrodynamic simulations presented in this
paper clearly show that, within the limitations of the idealized 
model, tidally-induced spiral shocks can lead to significant
radial transport in accretion disks in binary systems.  
Our results agree qualitatively with previous numerical work;
namely, we find that tidal forces excite a two-armed spiral shock
wave that remains stationary in the rotating frame of the binary
system.  However, our use of high spatial resolution and a high-order
numerical algorithm that conserves angular momentum about the
accreting star allows us to quantitatively measure the transport
effects of these spiral waves.

Perhaps the most interesting result is that $\alpha$ is large.
In all of
our simulations, the effective $\alpha$ was at least 0.1 in the
outer regions of the accretion disk where the tidal shock waves
are excited.  However, as predicted by analytic arguments, $\alpha$
decays with decreasing radius.  In cold disks such as those found
in CV systems, this drop is so precipitous that we could not measure
an $\alpha$ below a radius of only 1/3 that of the outer edge of the disk.

These numerical results agree in many respects with the analytic 
work of \citeasnoun{larson} and \citeasnoun{spru87}.  The shape of the 
spiral wave agrees reasonably well with the simple prediction that 
the trailing angle should be roughly the inverse of the orbital 
Mach number \cite{godon}, and the Mach number of the spiral shocks
matches the values found in the self-similar models of \citeasnoun{spru87}.
Even the value of the effective $\alpha$ in terms of the Mach number
of the spiral shocks, $M_s$, agrees with the analytic estimate
of \citeasnoun{larson}.

However, the value of the effective $\alpha$ in 
our simulations is much larger than the analytic predictions
of \citeasnoun{larson} and \citeasnoun{spru87} when written in 
terms of the orbital Mach number of the disk.  In addition, 
the dependence of $\alpha$ on sound speed and disk radius does not
agree with their analytic theory, which predicts an effective 
$\alpha$ proportional to  $(c_s^2 r)^{3/4}$.  This discrepancy 
thus appears to lie in the derivation of the Mach number of
the spiral shock as a function of the sound speed in the disk.

We note, however, that the analytic work applies to the
propagation of spiral waves through a disk, and does not give reference
to the source of the waves.  This work shows that tidal forces drive a
strong two-armed spiral wave, which in turn produces an effective
$\alpha \approx 0.1$ in the outer regions of the disk.  As this wave
propagates inward, there is a competition between growth and
dissipation, as well as the effects of an increasing Mach number.  In
our models dissipation wins out and the wave weakens (as measured by
radial velocity, density compression, or effective $\alpha$) as it
propagates to smaller radii.  Perhaps it would eventually approach the
analytic solution as it continues to weaken, but the corresponding
$\alpha$  of such a wave is much smaller than we would be able to
measure.

A further result of these simulations is the relative independence of
spiral waves on the mass ratio in the binary system.  The strength of
the two-armed spiral shocks at their origin near the outer edge of
the disk was fairly constant in all of our models.  This has the
consequence that, despite all else, one can be confident that the
effective $\alpha$ in the outer regions of an accretion disk in 
a binary system is (at least) $\sim 0.1$.

Finally, these models allow us to investigate the effects of tidal
truncation.  We found a strong dependence of the tidal radius on 
the Mach number in the disk, such that colder disks extended out
to larger radii.  The tidal radius in our coldest model (with $q=1$)
agrees well with the analysis of \citeasnoun{paczyn}, who found
the largest non-intersecting periodic orbit in a pressure-less disk.
Our results suggest that pressure effects in a warm disk can significantly
decrease the tidal radius (30\% smaller for a Mach number of $\sim 10$
at the outer edge of the disk).

There are three important caveats concerning our idealized model of
an accretion disk.  The first is our assumption of an isothermal
equation of state.  Internal heating in a real accretion disk will 
lead to a sound speed that increases with decreasing radii.  Spiral
waves in such a disk will not be as hampered by the rapid increase
in Mach number that we found in our simulations, so we might expect
the transport due to such waves to be significant deeper into the disk.
A real equation of state (with $\gamma > 1$) is also expected to produce
stronger transport effects \cite{spru87}.  Our second major assumption
is the limitation to two dimensions.  In general, one might expect the
inclusion of the third  dimension to result in weaker spiral waves
in the equatorial plane, since the vertical direction provides for  
the possibility of additional wave modes, vertical expansion that may
diminish the effects of shock compression, or refraction of waves 
out of the equatorial plane.

The third caveat is the missing tidal stream in our model.  We found
the most important effects of the spiral shocks to occur at the outer 
edge of the accretion disk, yet for a steady-state accretion disk
there must be some source of incoming gas.  In a close binary system
this is most likely a tidal stream originating at the inner Lagrange
point of the binary potential.  This stream will presumably strike
the outer edge of the disk and drive its own spiral wave into the
disk \cite{shu}.  Will this impact wave transport angular momentum?  
Will it interfere with the tidally-driven spiral waves?

While there is still a great deal of work to be done, it is clear
from this and previous work that tidally-driven spiral waves can
effectively drive mass transport through the outer regions of
an accretion disk in a close binary system.

\ack{}

This research was supported in part by an NSF Career grant and 
a Cottrell Scholars Award from the Research Corporation.
The numerical simulations described in this paper were computed on 
a Cray T916 at the North Carolina Supercomputing Center.

\end{document}